# Is a Direct Numerical Simulation of Chaos or Turbulence Possible: A Study of a Model Non-Linearity

Lun-Shin Yao

Department of Mechanical and Aerospace Engineering
Arizona State University
Tempe, Arizona 85287

Abstract

There are many subtle issues associated with solving the Navier-Stokes equations. In this paper, several of these issues, which have been observed previously in research involving the Navier-Stokes equations, are studied within the framework of the one-dimensional Kuramoto-Sivashinsky (KS) equation, a model nonlinear partial-differential equation. This alternative approach is expected to more easily expose major points and hopefully identify open questions that are related to the Navier-Stokes equations. In particular, four interesting issues are discussed. The first is related to the difficulty in defining regions of linear stability and instability for a time-dependent governing parameter; this is equivalent to a time-dependent base flow for the Navier-Stokes equations. The next two issues are consequences of nonlinear interactions. These include the evolution of the solution by exciting its harmonics or sub-harmonics (Fourier components) *simultaneously* in the presence of a constant or a time-dependent governing parameter; and the sensitivity of the long-time solution to initial conditions. The final issue is concerned with the lack of convergent numerical chaotic solutions, an issue that has not been previously studied for the Navier-Stokes equations. The last two issues, consequences of nonlinear interactions, are not commonly known. Conclusions and questions uncovered by the numerical results are discussed. The reasons behind each issue are provided with the expectation that they will stimulate interest in further study.

## 1. Introduction.

It is well known that many difficult phenomena associated with the Navier-Stokes equations are subtle and subject to different interpretations. In this regard, four issues are of interest here. The first is a linear-stability analysis for a time-dependent governing parameter; the next two are related to nonlinear interactions; and the last is concerned with the lack of convergent numerical chaotic solutions. The understanding of these issues is not straightforward since no general analytical solutions exist. The purpose of this paper is to communicate the results of some recent progress on these issues and other related matters that are the results of a study of a model partial different equation, the one-dimensional Kuramoto-Sivashinsky (KS) equation. Hopefully, this simple model equation will help to clarify the role of nonlinear terms and their consequences.

It is noteworthy to point out that two most important facts among many issues about non-linear differential equations, which not commonly known, are:

1. No computed chaotic solution, which is independent of the integration time-step employed, exists.

2. A sensitivity-to-initial conditions is generally viewed as a *necessary* and *sufficient* condition for the existence of chaos. However, this property is also noted in the solutions of all non-linear differential equations when the value of their governing parameters is larger than their critical value, [2, 12-14]. Consequently, it cannot be argued that this sensitivity is a sufficient condition for chaos.

The spectral method used to solve the KS equation is described in section 2. In section 3, it is shown that a linear-stability analysis for an unsteady governing parameter can be carried out by a direct numerical integration of the linearized differential equations to find its critical state, the boundary between stable and unstable states. Such an approach is equivalent to the classical linear-stability analysis, which calculates eigenvalues when the parameter is a constant. In a benchmark paper, Hall [2] used this direct method to analyze the linear spatial instability of developing Taylor-Görtler vortices.

The next two issues involve the behaviors of solutions when the value of the governing parameter is above the critical state; in this case, nonlinear interactions among disturbances of non-zero amplitudes become important. Such behaviors are the key factors in the understanding of various transition processes in fluid flows. Classical treatments of nonlinear interactions can be traced to the pioneering efforts of Landau [5], Stuart [9], and Watson [10] for shear flows, and Philips [6] for water waves. In section 4a, it is shown that a linearly stable or unstable initial condition always excites the entire set of harmonics appearing *simultaneously* in its Fourier components, no matter how small their amplitudes. Consequently, at the critical state, the solution is nonlinearly stable since the influence of the linear term vanishes for the critical wave, which can transfer energy simultaneously to its dissipative harmonics. Another natural consequence is that a *single-wave* (Fourier component) solution does not exist for a non-linear partial differential equation since the entire set of harmonics is excited. This result provides a simple explanation as to why a long wave can trigger many short waves simultaneously, and why energy is not necessarily transferred from long waves to short waves in cascade.

It has been previously demonstrated [13, 14] that the nonlinear terms of the Navier-Stokes equations can be interpreted as forced and resonant vibrations, and can induce "energy" transfer among waves. Here energy is expressed by the square of the wave amplitude. These ideas appear originally in the pioneering work of Phillips [6] for water waves. Since forced vibrations do not cause significant transfer of energy among individual waves, they do not change the spatial structure of the solution. However, they do excite all wave harmonics, thereby providing a starting point for resonant energy transfers. In contrast, the energy transferred due to resonance is substantial, and can lead to observable changes in the solution structure. Consequently, the energy transfer associated with the nonlinear terms cannot be ignored as long as wave amplitudes are not zero. Examples are presented in sections 4a and 4b to illustrate the existence of such behaviors for the KS equations when the governing parameter is constant (4a) or time-dependent (4b). These numerical results also show that the energy can be transferred to sub-harmonics.

In section 4c, examples are used to show that the long-time solutions of the KS equation at a fixed parameter value are sensitive to initial conditions. Previous work has shown that such behavior exists for the Navier-Stokes equations [2, 12, 13]. This suggests that the classical principle of "dynamic similarity" is strictly valid only if the governing parameter is below its critical value.

The final issue involves the non-convergence of numerical computations for values of the parameters that allow chaotic solutions for the KS equation. Rössler [7] first noted this issue while commenting on the numerical approximation of the chaotic solution of the celebrated Rössler system of three ordinary differential equations. In section 5, it is shown that no convergent, long-time numerical chaotic solution can be found by all time-discrete methods and time steps tested in this study. The lack of *convergent* computational results for chaos has also been found for the Lorenz system [15]. This leads to an interesting question: Is it possible to obtain a *convergent* direct numerical turbulent solution of the much more complicated Navier-Stokes equations?

## 2. Analysis

A spectral method is used to generate a spatially discrete form of the KS equation

$$u_t + 4u_{xxxx} + \lambda\left(u_{xx} + uu_x\right) = 0, \tag{1}$$

for x in the interval [0,2π], where λ is the governing parameter that measures the energy production and the importance of the nonlinear term. An approximate solution can be expressed as

$$\mathrm{u(x,t)} = \sum_{\mathrm{n=1}}^{\mathrm{N}}\left(\mathrm{A_n(t)\,e^{inx} + A_{-n}(t)\,e^{-inx}}\right) , \tag{2}$$

where $A_{-n}$ is the complex conjugate of $A_n$. Each Fourier component of (2) is interpreted as a wave, and $A_n$ is the complex amplitude of the wave. The energy of each wave is represented by the square of its amplitude, and the total energy is the sum of these individual wave energies. Since the wave corresponding to n = 0 does not interact with any others, as is common, it is not included in this discussion.

In order to demonstrate the structure of the resulting discrete system of ordinary differential equations, this system of equations is displayed as an example for N = 5:

$$\frac{d}{dt}\begin{bmatrix} A_1 \\ A_2 \\ A_3 \\ A_4 \\ A_5 \end{bmatrix} = \begin{bmatrix} 1^2\left(\lambda - 4\bullet 1^2\right) & & & & \\ & 2^2\left(\lambda - 4\bullet 2^2\right) & & & \\ & & \bullet & & \\ & & & \bullet & \\ & & & & 5^2\left(\lambda - 4\bullet 5^2\right) \end{bmatrix}\begin{bmatrix} A_1 \\ A_2 \\ A_3 \\ A_4 \\ A_5 \end{bmatrix} - i\lambda\begin{bmatrix} 0 & A_{-1} & A_{-2} & A_{-3} & A_{-4} \\ A_1 & 0 & 2A_{-1} & 2A_{-2} & 2A_{-3} \\ A_2 & 2A_1 & 0 & 3A_{-1} & 3A_{-2} \\ A_3 & 2A_2 & 3A_1 & 0 & 4A_{-1} \\ A_4 & 2A_3 & 3A_2 & 4A_1 & 0 \end{bmatrix}\begin{bmatrix} A_1 \\ A_2 \\ A_3 \\ A_4 \\ A_5 \end{bmatrix}. \tag{3}$$

This system shows that energy transfers can occur from small to large wave numbers (small to large n) in sequence, which is an energy cascade. However, this cascade is only one of many possible nonlinear energy-transfer mechanisms; moreover, it is not necessarily the dominant one. This conclusion is also valid for the Navier-Stokes equations since they take on a similar structure to (3) when expressed in terms of a suitable problem-dependent version of the expansion in (2) [14].

A large number of computations for (3) have demonstrated that energy transfers usually occur from high-amplitude waves to low-amplitude waves, but the transfers do not always follow this “rule.” Otherwise, this discrete system is similar to the discrete system for the amplitude-density functions of a Fourier-eigenfunction spectral method for the Navier-Stokes equations [2, 12, 13], which suggests that the solution properties studied in this paper are relevant to those of the Navier-Stokes equations.

Two different time-integration methods are used for the temporal discretization of (3). The first method, which will be referred to as the explicit method, approximates the non-linear terms by an explicit second-order-accurate Adams-Bashforth method and the linear terms by an implicit Crank-Nicholson method. The second method, which will be referred to as the implicit method, uses a second-order Crank-Nicholson method for both nonlinear and linear terms. A careful comparison of many numerical computations show that N = 32 is sufficient for the examples of this paper.

## 3. Linear Stability Analysis

Like in the well-studied linear hydrodynamic stability, the linear terms of (3) represent the difference between energy production and dissipation. If this difference has a positive value, then a mode grows after being excited and is classified as linearly unstable. Ignoring the nonlinear terms in (3), the amplitude functions $A_n$ can be easily determined. This leads to the following conclusion. The linear-stability boundary can be determined by the condition $dA_n/dt = 0$, even with $\lambda$ time-dependent. This states that a wave is linearly unstable when the time derivative of its amplitude is positive or linearly stable when it is negative. This is consistent with the traditional linear stability analysis of hydrodynamics of steady bases flows that computes eigenvalues. The advantage of directly solving an initial-value problem to determine the sign of $dA_n/dt$ is that this approach is not limited to problems with a constant λ or a steady base flow. Its weakness is that it may require assuming initial values for the dependent variables in order to carry out the computation for more complex equations such as the Navier-Stokes equations. Hall [2] used this direct method to analyze the linear spatial instability of developing Taylor-Görtler vortices, and resolved several controversies. For the KS equation, the linear-stability analysis is trivial, and the stability boundary can be determined by either an eigenvalue method or a direct method: $A_n$ is neutrally stable at $\lambda = 4n^2$.

## 4. Nonlinear Stability Analysis

### 4a. Constant λ

The first nonlinear case considered uses $\lambda = 4$, for which the wave $A_1$ is neutrally stable according to a linear-stability analysis. The initial condition is $A_1 = 1$, with all other amplitudes zero. The time step is 0.001 and N=32. It has been shown that the non-linear terms represent many wave resonances and are not just limited to resonant trios [12]. Results obtained with the implicit method showing that the energy transfers from $A_1$ to its harmonics occur *simultaneously* are displayed in Figure 1. The energy transfer from $A_1$ causes the initial growth of all harmonics. All waves start to slowly decay after 60 time steps, as shown in Figure 2. Consequently, the critical wave is not neutrally stable at the critical point if non-linear effects are considered.

On the other hand, results obtained by the explicit method show an energy cascade in Figure 1. In agreement with the resonance conditions associated with nonlinear wave-interactions [6, 15], at the first time step $A_2$ starts to grow; at the second time step, $A_3$ and $A_4$ are excited, and so forth. These numerical results are incorrect initially because they are only functions of the number of integration time-steps, and not of the time, as shown in Figure 1. The initially incorrect results of the explicit method eventually coincide with the convergent ones of the implicit method. This time delay in transferring energy from small to large wave numbers can be traced to the limitations of the explicit method for nonlinear differential equations. Comparing the results of the computations for the explicit and implicit methods shows that there is a small time delay in the response of short waves as predicted by the explicit method in the presence of an initial condition that favors long waves. Thus, an implicit scheme may be more attractive for unsteady nonlinear problems.

The next example presented in Figure 3 shows that energy can transfer to sub-harmonics when the resonant conditions are satisfied. The value of $\lambda$ is 4, the initial conditions are $A_2 = A_3 = 1$, with all others zero. The computational time-step is set at $\Delta t = 10^{-5}$ and N = 40 in order to show clearly its quick development for small time. The results show that $A_1$ gains energy initially from its harmonics since it is neutrally stable at $\lambda = 4$.

**4b. Time-Dependent λ**

For cases in which $\lambda$ is a function of time, solutions may reflect a wide range of the governing parameter. In the following examples, $\lambda = 40t$ is selected, with N=50 and $\Delta t = 5x10^{-6}$, in order to obtain convergent results. The results for an initial amplitude $A_1 = 1$, with all other amplitudes zero, are given in Figure 4. Initially, the amplitude of $A_1$ decreases because this mode is linearly neutrally stable until $\lambda$ becomes larger than 4. The important non-linear effect is that $A_1$ produces and transfers energy to its entire range of harmonics, demonstrating that these effects can generate harmonics almost instantaneously. The initial amplitudes of the harmonics of $A_1$ are very small, but, when they become unstable, they grow quickly as indicated by the linear terms of equation (3). Different waves become dominant as the value of $\lambda$ increases in time.

After t = 4, the computed results becomes chaotic, and no convergent computational results can be obtained. This topic is the subject of the next section. It is interesting to note that the computed result becomes non-chaotic again after t ~ 5. This agrees with previous studies showing that the KS equation has several narrow windows for $\lambda$ that allow chaos [4].

The second unsteady case has $A_2 = 1$, with all other amplitudes zero, at t = 0. The results are plotted in figure 5. Initially $A_2$ decreases, but then increases when it becomes unstable according to the linear-stability analysis as $\lambda$ increases beyond 16 at t = 0.4. The initial energy transfers from $A_2$ to $A_4$, $A_6$, etc. occur simultaneously. After t = 0.4, the amplitudes of all waves start to increase due to the fact that $A_2$ becomes unstable and has more energy, which can be transferred to other waves. This is why the evolution of the amplitudes of all even waves is almost the same as the evolution of $A_2$ initially. No odd waves are excited in this example since they are not harmonics of $A_2$ and their initial amplitudes are zero.

These cases demonstrate an important non-linear effect showing that any disturbance can produce its entire range of harmonics. This is a fast *seeding* process and does not occur slowly in sequence. As a result, some of the induced waves may become unstable and lead to a transition from stable behavior to chaos. Unfortunately, no convergent computed results can be found once transition starts; this topic will be addressed in the next section. Since a wave can simultaneously transfer energy to its entire set of harmonics, a long disturbance wave can trigger short waves. The well-known Tollmien-Schlichting waves of boundary-layer instability are an example of such a natural non-linear process.

**4c. Sensitivity to Initial Conditions**

Another important observation already apparent in Figures 4 and 5 is that computed results are *sensitive to initial conditions*. To pursue this property more fully, two cases with a fixed value of $\lambda = 17$ are plotted in Figures 6 and 7 to demonstrate the existence of multiple equilibrium solutions. The integration time-step is set at $10^{-4}$ and N=64. When $\lambda = 17$, the first two waves are unstable. In Figure 6, the initial condition is $A_1 = 0$, $A_2 = 0.1$, with all remaining waves having amplitudes set at $10^{-5}$. Since $A_3$ is stable and its initial amplitude is $10^{-5}$, energy is transferred to it via resonance. After $A_3$ gains energy, it can resonant with $A_2$ and $A_3$ to transfer energy to $A_1$. This is an example of reverse energy transfer to subharmonics. If the initial condition is changed to $A_2 = 0.1$, with all others zero, then only even waves will be excited.

The case plotted in Figure 7 has a slightly different initial condition ($A_1 = 10^{-5}$) from the case plotted in Figure 6. Since the initial amplitude of $A_1$ is not exactly zero, $A_1$ gains energy linearly because it is unstable at $\lambda = 17$. These two figures clearly indicate that slightly different initial conditions can lead to different equilibrium solutions. This shows that stable long-time solutions can be sensitive to initial conditions without necessarily being chaotic solutions. These results imply the existence of different *domains of attraction*. The starting point of a computation within a particular domain of attraction determines the particular position of the final numerical solution. Testing many cases with slightly different initial conditions leads to the conclusion that infinitely many long-time periodic solutions can exist for a given $\lambda$. Consequently, a rich variety of long-time solutions can be expected.

The existence of multiple long-time flows is widely recognized for certain problems in fluid mechanics. Coles [1] and Snyder [8], in benchmark papers, demonstrated that these flows depend on the initial conditions for a given final Taylor number in Taylor-Couette flows. The essence of their observations has been confirmed by recent direct numerical solutions of the Navier-Stokes equations [2, 12]. Similar numerical results are available for the totally different physical example of mixed convection in a vertical annulus [13]. This work establishes the possibility that a instability can lead to multiple solutions. If this is so, initial conditions might be additional factors to be considered in the application of the principle of *dynamic similarity*.

The question of extending this conclusion to turbulence lacks consensus. Indeed, this is a very difficult question to answer experimentally. Unfortunately, its answer is also beyond the scope of current numerical methods. The reason is described in the next section.

## 5. Chaos

It has been shown that the numerical solution of the KS equation is chaotic for $\lambda = 69$ [4]. The results presented in this section are computed with N = 64. A computation with the initial condition of a single non-zero wave, with the remaining waves at zero, indicates that the numerical solution is not chaotic. If the initial amplitudes of the linearly unstable waves corresponding to n = 1 to 4 are set to a small number, then the numerical solution becomes chaotic. Once such computed solutions become chaotic, they are time-step sensitive and break down after a very short time, about $t \geq 0.41$. The results presented in Figures 8 and 9 are obtained with the initial amplitude of all waves set at $10^{-5}$. Since the difference between the production and dissipation for $A_3$ has a maximum value at $\lambda = 69$ and its shape is representative, only $A_3$ is plotted.

The time history of $A_3$ obtained by the explicit method is plotted in Figure 8 for five different values of $\Delta t$. The numerical result grows without bound for $\Delta t > 0.0007$. Reduction of the time step to $10^{-4}$ extends the convergent solution from t = 0.2 to t = 0.4. Further reduction of $\Delta t$ to a value smaller than $10^{-4}$ does not improve the results. The solution becomes time-step sensitive after t = 0.4. Several different numerical methods (Adam-Bashforth methods up to the fifth order; the implicit method mentioned in section 3; fourth-order Runge-Kutta method; and other methods involving variable time steps) did not yield a convergent solution.

A plot of the time-averaged energy spectrum

$$E(n=3) = \frac{1}{t}\int_0^t \left|A_n^2\right| dt \; = \frac{1}{M}\sum_{m=1}^{M} A_n(m \bullet \Delta t) \bullet A_{-n}(m \bullet \Delta t),$$

in Figure 9 shows that the time-averaged solutions also depend on the time-step. The case labeled FFT in Figure 9 is obtained by a pseudo-spectral method. This spectrum is an important statistical property for chaos, and cannot be a function of the integration time-step used in obtaining it; otherwise the integration time-step becomes an additional artificial problem parameter without a physical meaning. The lack of convergence in the results of Figures 8 and 9 is, at first glance, unexpected, but is real. Attempts to ignore this behavior frequently rely on three commonly believed, but erroneous arguments. However, these arguments, which are stated below, cannot withstand careful scrutiny.

*Argument 1*: Since a necessary property of chaos is the presence of a positive Liapunov exponent, or a positive nonlinear exponential growth-rate, the truncation error introduced by various numerical methods can be amplified exponentially. Hence, erroneous solutions develop differently due to different truncation errors. This is equivalent to saying that the finite-difference equations, which approximate the differential equations, are unstable. Thus, since convergence requires stability and consistency, convergent computed results are not achievable. However, such unstable cases are shadowable, that is, they remain sufficiently close to the true trajectory with slightly different initial conditions.

However, Argument 1 is not valid uniformly in the entire geometric space as demonstrated in [10] and [15]. The breakdown in the numerical solutions for chaos shown in Figures 8 and 9 is sudden, explosive, and unshadowable, but it is not due to the exponential growth of numerical errors associated with an unstable manifold.

*Argument 2*: It is well known that chaotic solutions of differential equations are sensitive to initial conditions. The different truncation errors associated with different integration time-steps, in effect, lead to a series of modified initial conditions for later times. Consequently, computed chaotic solutions are integration time-step dependent, and cannot be considered to be an approximate, in any sense, solution of the differential equations.

On the other hand, as demonstrated in this paper, stable long-time numerical solutions for the one-dimensional KS equation are sensitive to initial conditions, but are also convergent and independent of the integration time-steps. This shows that a solution sensitive to initial condition is not necessarily sensitive to integration time-steps. A commonly cited computational example in chaos involves two solutions of slightly different initial conditions that remain “close” for some time interval and then diverge suddenly. In fact, this behavior is often believed to be a characteristic of chaos. More properly, this phenomenon is actually due to the explosive amplification noted above.

*Argument 3*: It is commonly believed that the existence of an attractor (inertial manifold) guarantees the long-time correctness of numerical computations, irrespective of the numerical errors that are inevitably present in any computation of chaos. Such a concept has never been proved, but it is usually used to support the belief that numerical errors do not invalidate particular computed chaotic results among the community working on numerical solutions of dynamic systems.

This argument is incorrect because a computation contaminated by numerical error can escape an existing correct attractor. The existence of an attractor does not guarantee that a numerical computation of chaos, which is *unavoidably* contaminated by whatever numerical errors that exist, is acceptable. This has been demonstrated in Figure 9 of [15].

The results in [15] show that any numerical errors, no matter how small, in the numerical solution of the Lorenz system of differential equations can be amplified dramatically. This is due to the fact that these equations are not a “*hyperbolic*” system and have a finite number of singularities [10]. The inset of a singular point (a collection of points such that all trajectories starting at those points travel toward the singular point) forms a local (virtual) "separatrix" (a finite surface separating two regions in a geometric space). From a geometric viewpoint, any trajectory touching a virtual separatrix will approach the singular point asymptotically; thus, a solution trajectory cannot penetrate a separatrix. The breakdown of the numerical computations is a consequence of the fact that a small numerical error can force the computed trajectory to penetrate the virtual "*separatrix*," hence, violating the uniqueness theory for differential equations. This behavior can cause a dramatic amplification of numerical errors, can occur repeatedly, and can lead to an attractor whose shape differs substantially from one without numerical errors; see Figure 9 of [15]. The results of [15] demonstrate that a computed result, even though contaminated by numerical errors, is associated with an attractor. It is important to recognize that different computed results produce different attractors. An important conclusion of [15] is to expose the belief that the existence of a unique attractor is sufficient to ensure that computed results are always "correct," no matter what the numerical errors are.

## 6. Conclusions

It is noteworthy that the simple one-dimensional KS equation, as a model, contains considerable information closely related to many important problems that have been extensively studied in fluid dynamics. Since the required analysis for the KS equation is relatively simple, the computed results can be interpreted more readily than the more complex Navier-Stokes equations.

The conclusions of this paper are outlined in the following remarks, with reference to the four issues identified in section 1. It is significant to note that the conclusions for the first three issues are also valid, at least in some problems, for the Navier-Stokes equations:

1. The linear-stability boundary can be identified by a direct numerical integration of the linearized differential equation for either a steady or a time-dependent parameter.

2. A small disturbance can excite the entire set of Fourier components instantaneously so that, while the effect of nonlinear interactions may be small initially, they can determine the long-time evolution of the solutions.

3. There are multiple stable long-time solutions dependent on the imposed initial conditions beyond the onset of instability.

4. Current discrete numerical methods can compute solutions beyond the onset of instability very efficiently as long as they are not chaotic. This issue has not been addressed for the Navier-Stokes equations.

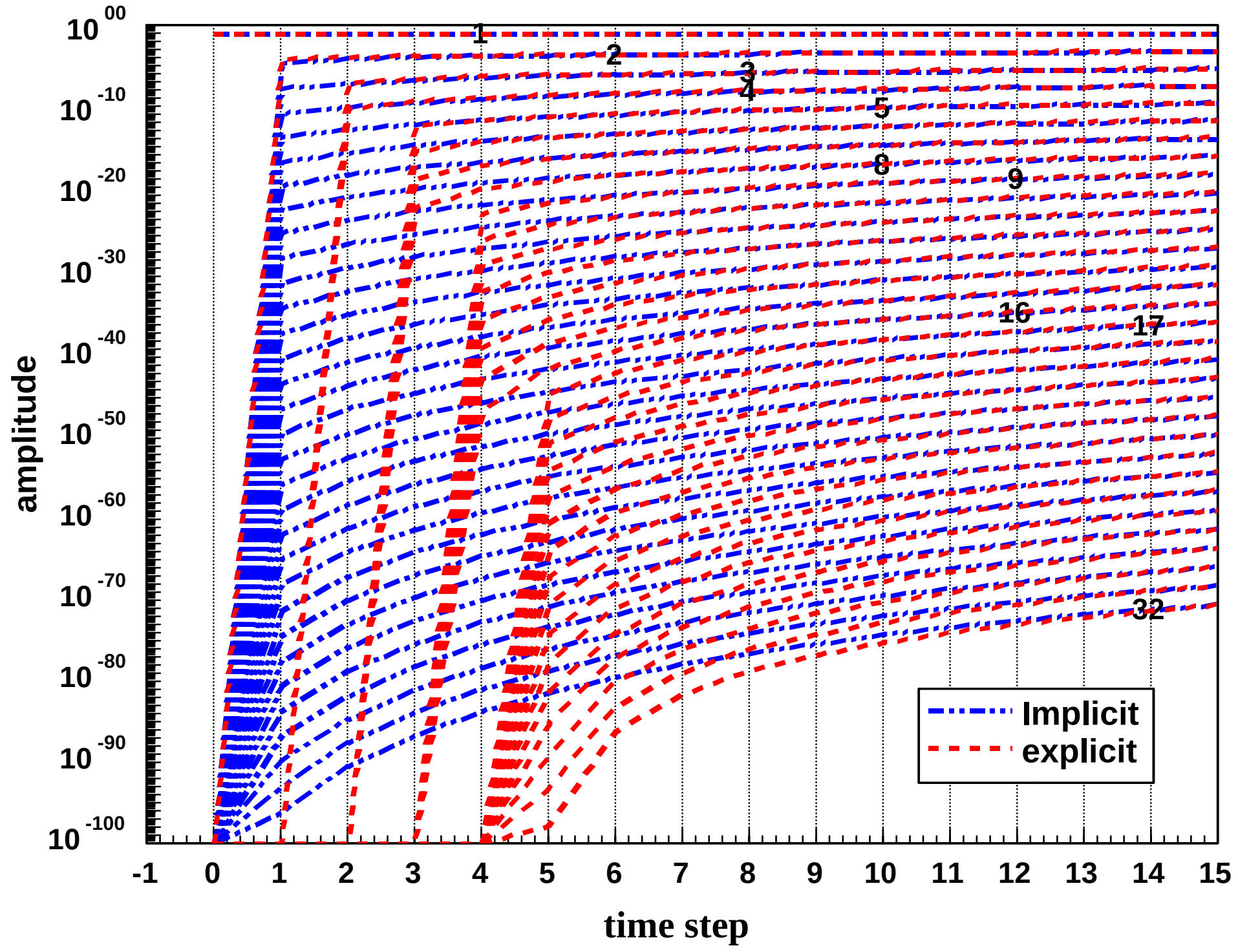

Figure 1. Non-linear evolution at the critical point for the initial condition that $A_1=1$.

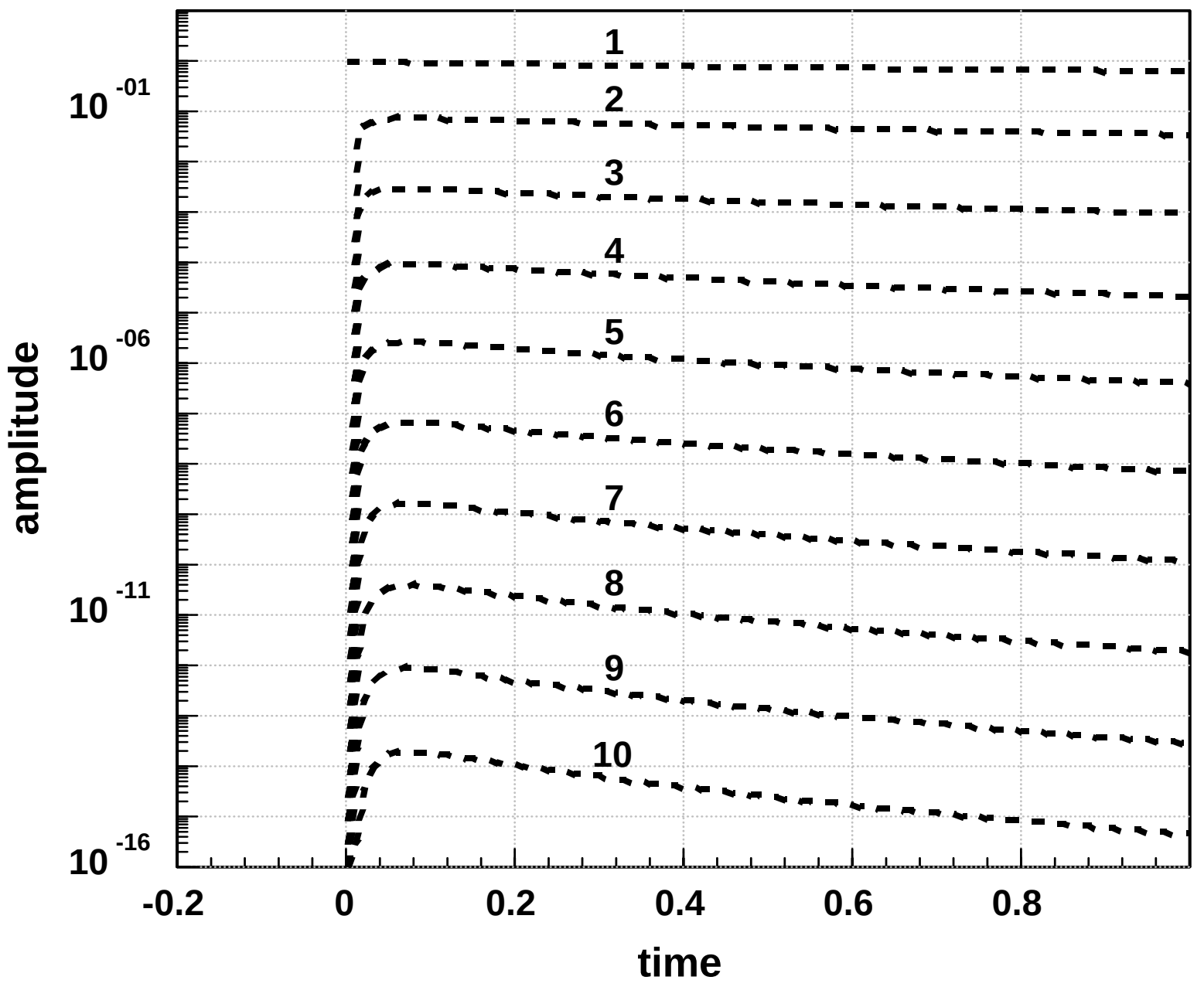

Figure 2. Non-linear evolution of the first ten waves at the critical point for the initial condition that $A_1=1$. $\Delta t=0.001$

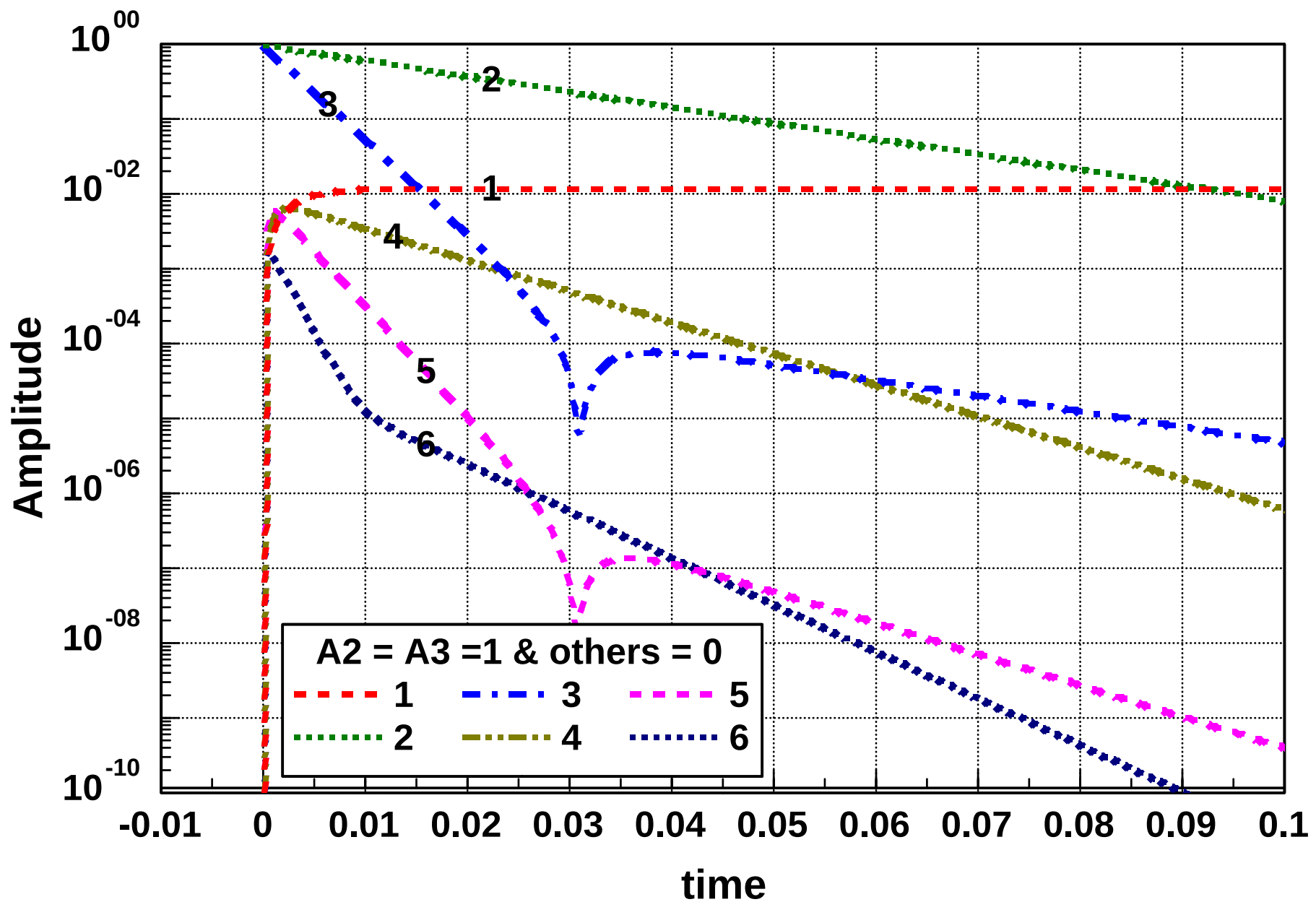

Figure 3. The value of λis 4, so $A_1$ is neutrally stable and other waves are stable. The initial condition is $A_2$ = $A_3$ =1 and other waves are zero. The results show $A_1$ gain energy from $A_2$ and $A_3$ initially. This is an example of reverse resonant energy transfer that energy is transferred to sub-harmonics when the resonant conditions are satisfied.

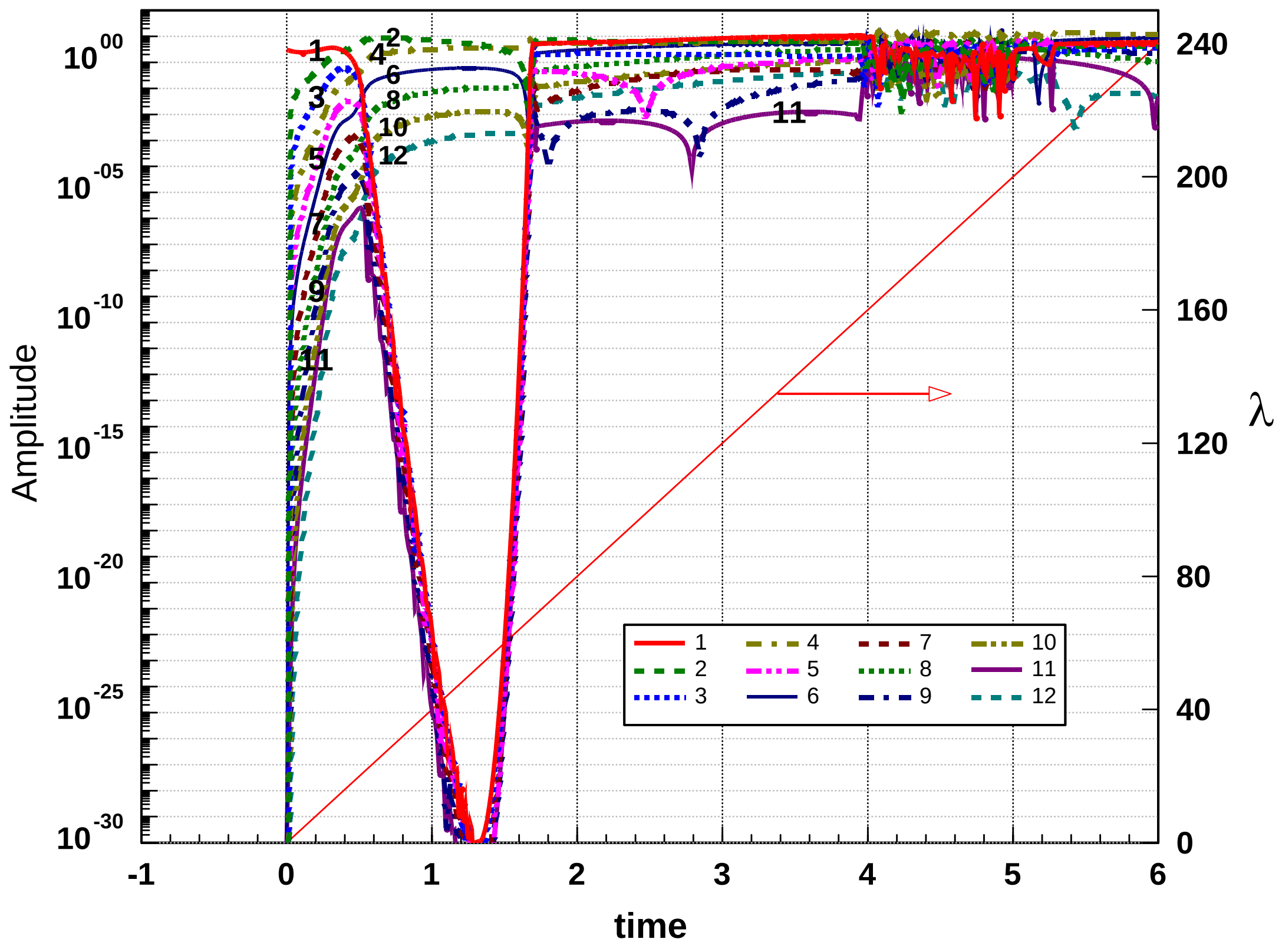

Figure 4. The initial amplitude of $A_1$=1 and $\lambda$=40t. The entire harmonics of $A_1$ are excited. The initial increase of $A_2$ is completely due to resonant energy transfer from $A_1$. The wave $A_1$ starts to decrease after $A_2$ becomes unstable at $\lambda$=16. This example shows that a long-wave disturbance can excite a full spectral of disturbances by non-linearly producing its harmonics. Only the first 12 waves are plotted.

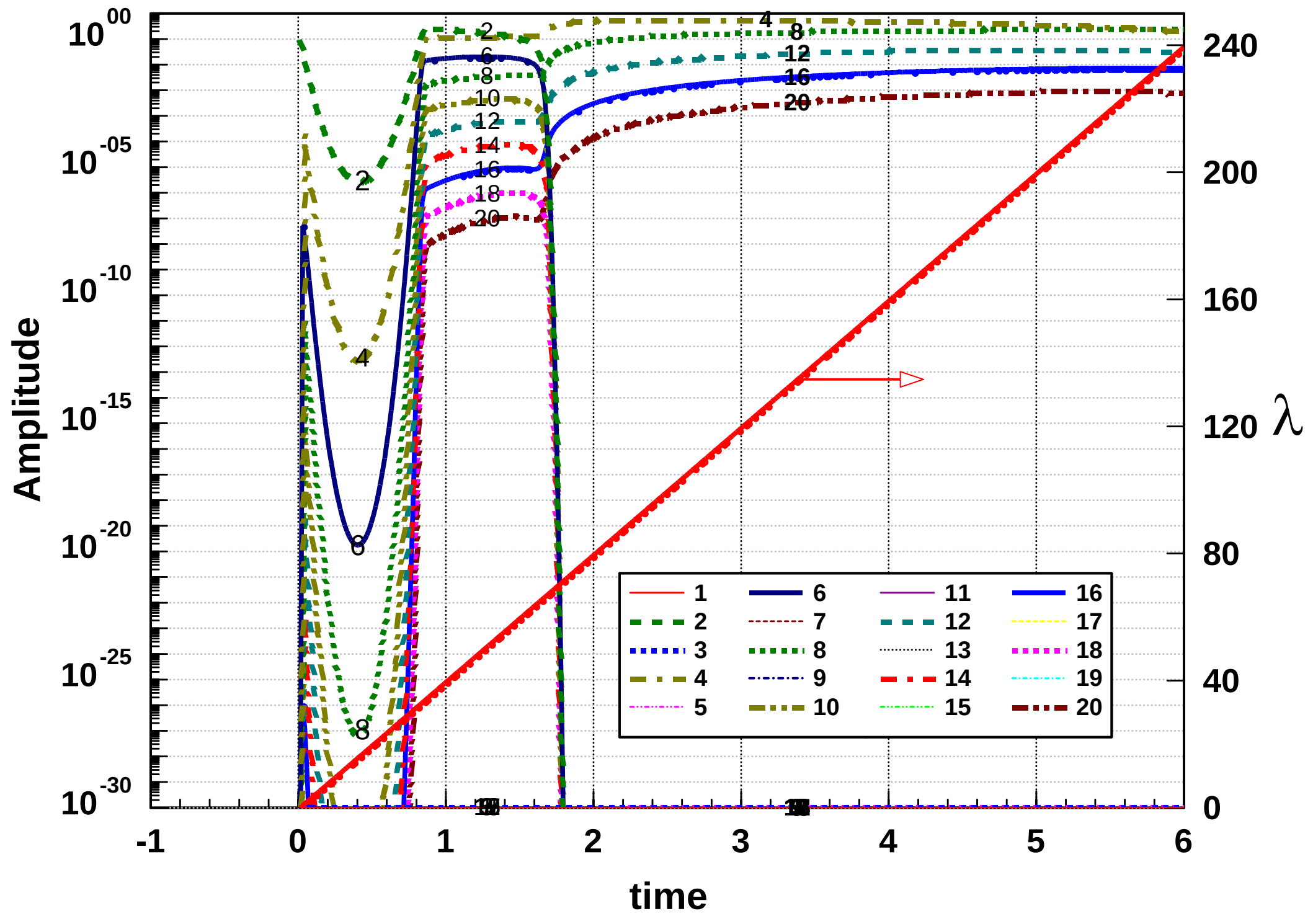

Figure 5. The initial amplitude of the wave $A_2$ =1 and λ=40t. Only the first 20 waves are plotted. No odd number waves are excited since their initial amplitudes are zero. All harmonics of $A_2$ grow almost simultaneously initially

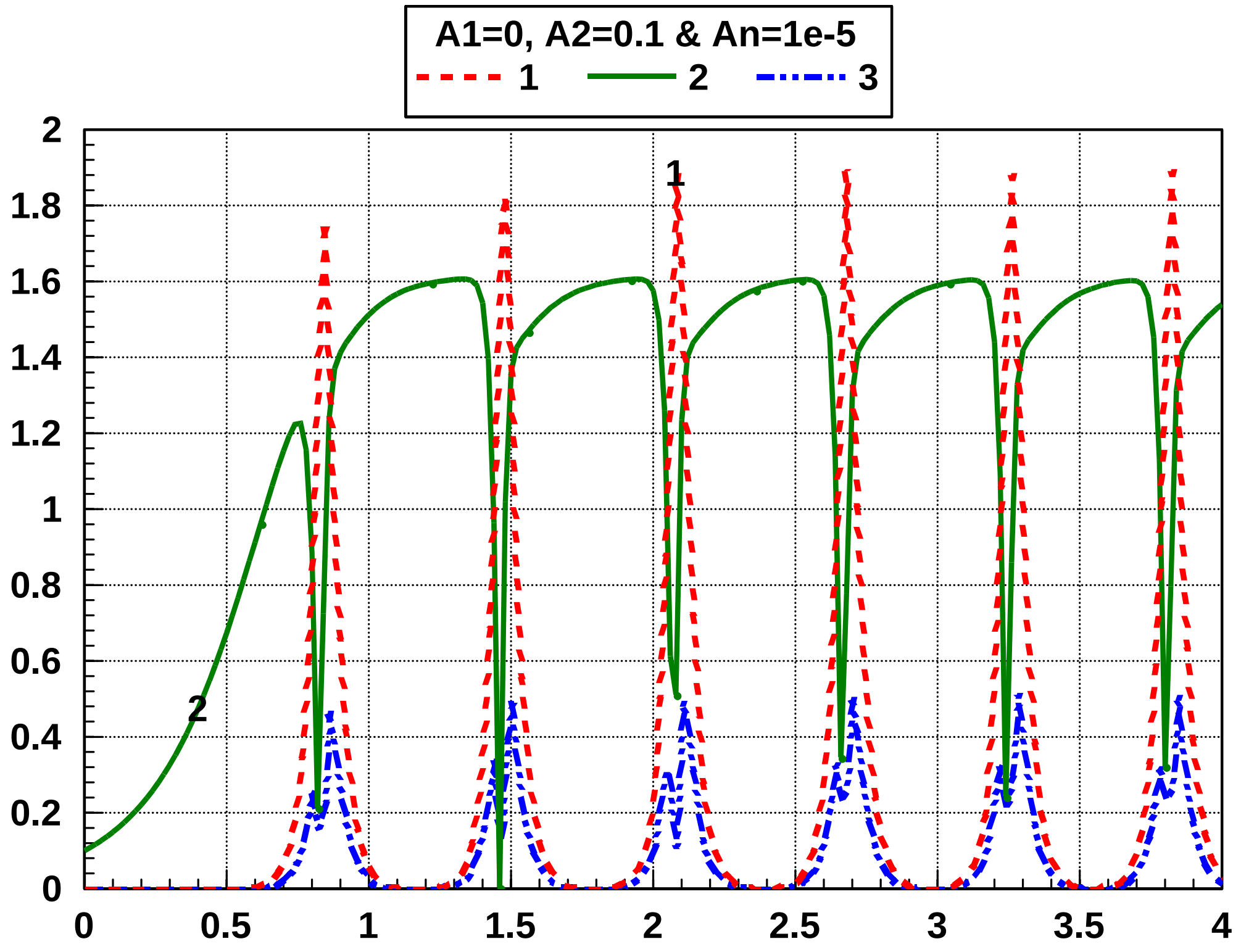

Figure 6. The value of λ is 17, so the first two waves are unstable. The initial amplitude of $A_1$=0, $A_2$=0.1 and the amplitude of the rest waves is set at $10^{-5}$. Since $A_3$ is stable and its initial amplitude is $10^{-5}$, the energy is transferred to it via resonance. Then $A_3$ transfer energy to $A_1$ resonantly. This is a reverse transfer to subharmonics. If the initial condition is $A_2$=0.1 and all other waves are zero, then only even waves will be excited.

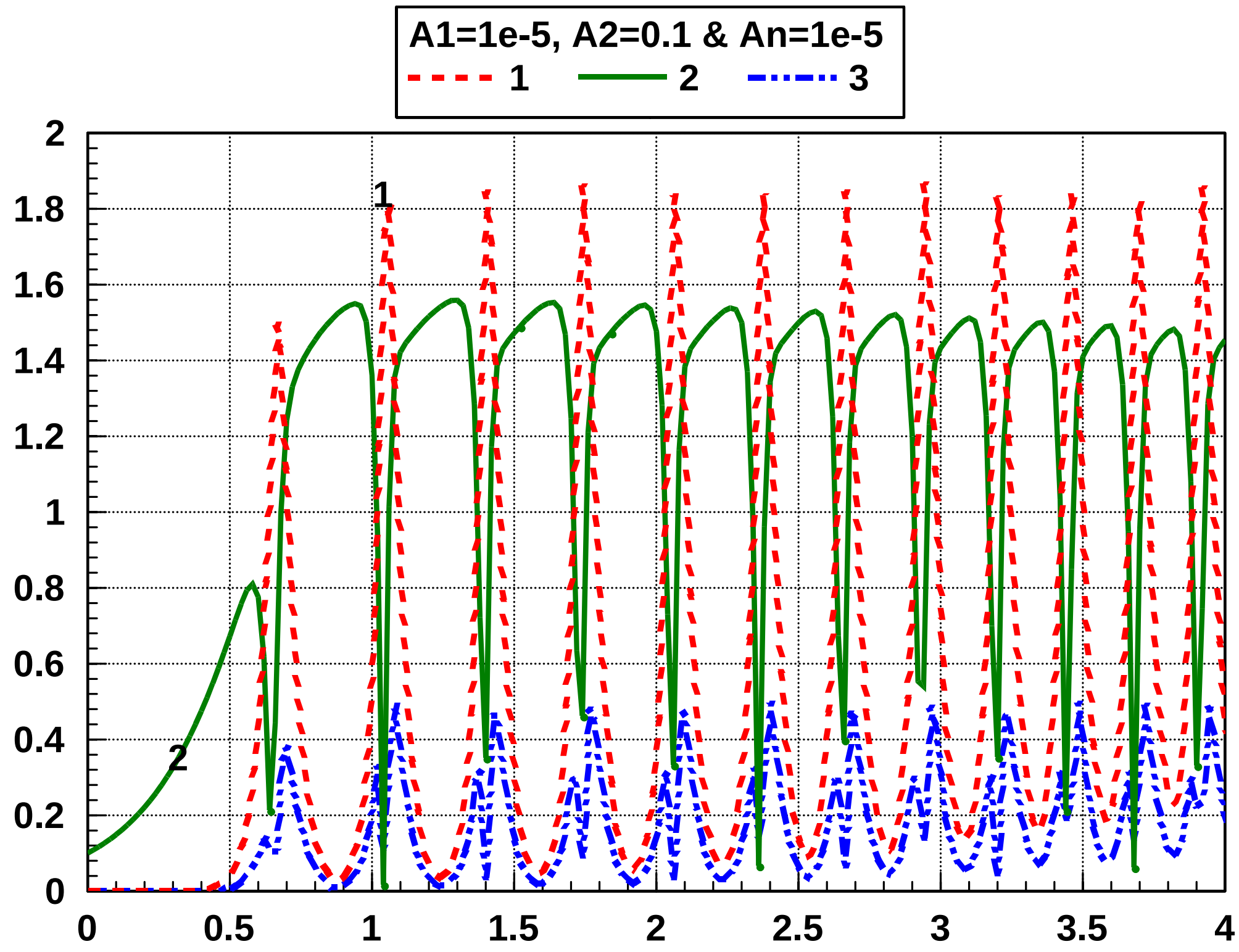

Figure 7. The value of λ is 17, so the first two waves are unstable. The initial amplitude of $A_2$=0.1 and the amplitude of the rest waves is set at $10^{-5}$. Since the initial amplitude of $A_1$= $10^{-5}$ is not exact zero, the increase of $A_1$ is due to it is unstable. The solutions of Figures 5 and 6 are obviously not same due small difference of the initial conditions in $A_1$, even though the λ values of two cases are same.

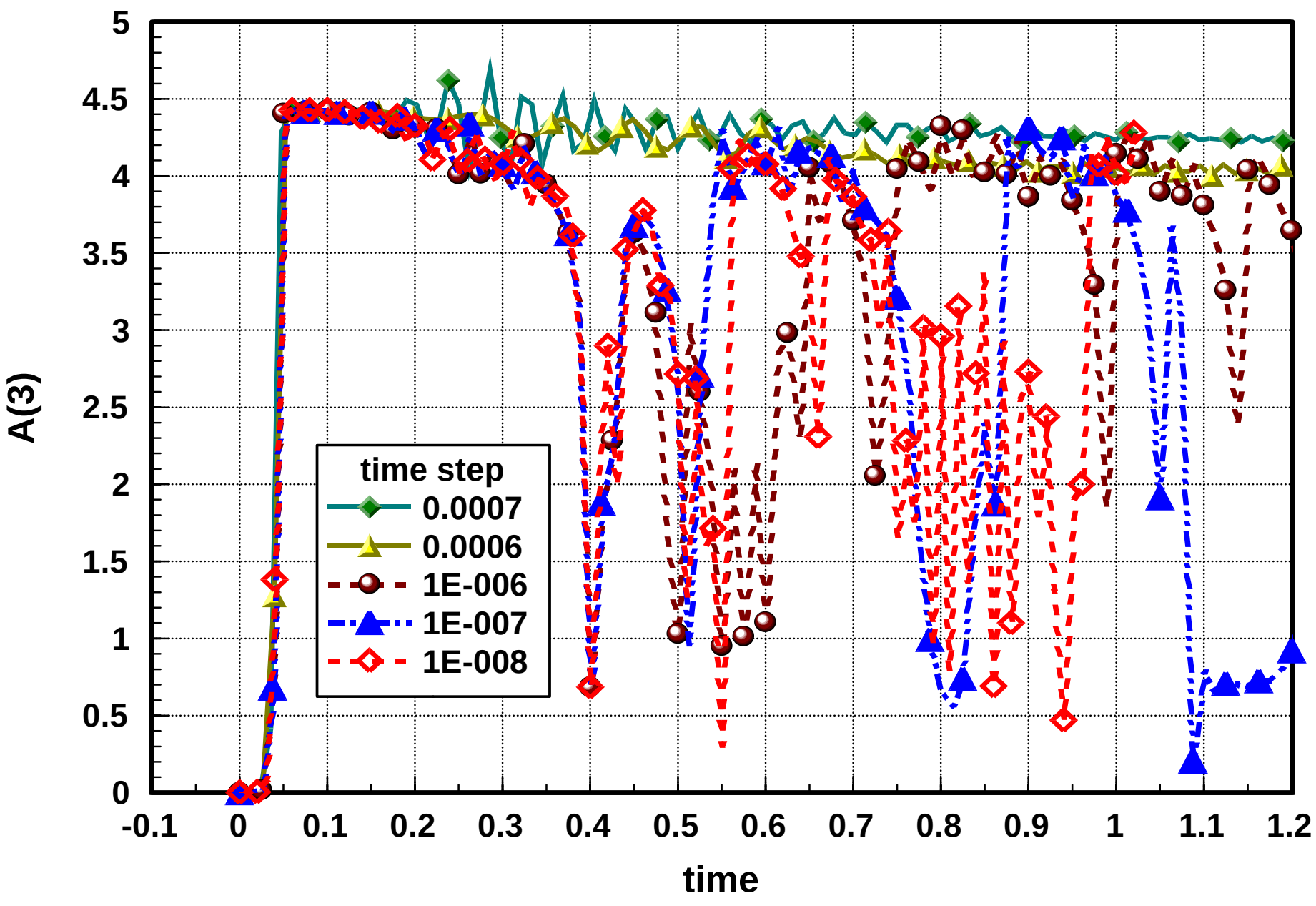

Figure 8. Time history of the amplitude of the third wave for various time steps

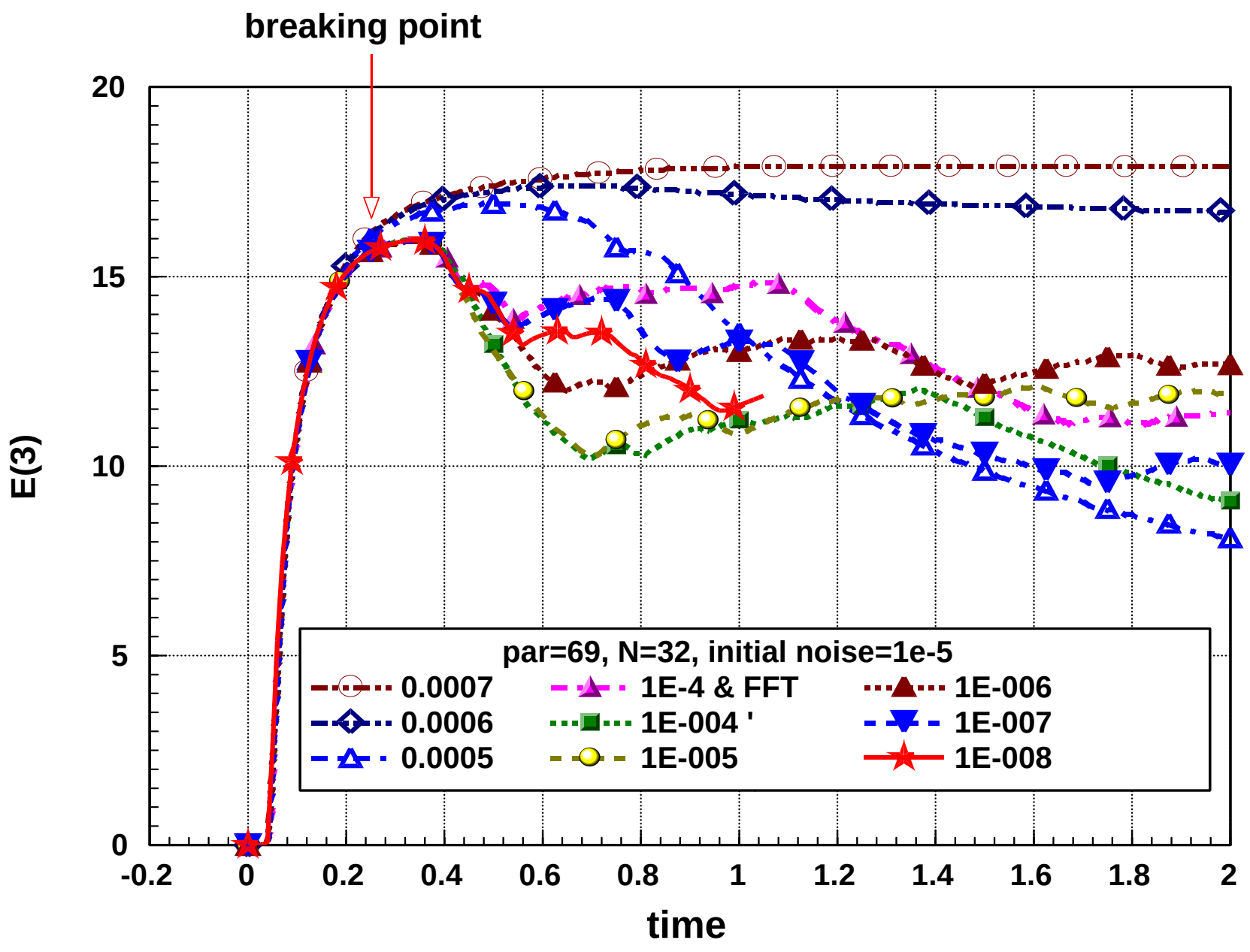

Figure 9. Time-averaged energy for the third wave for the various time steps